\documentclass[superscriptaddress,aps,prd,preprintnumbers,nofootinbib,preprint,12pt]{revtex4}
\usepackage{setspace}
\usepackage{amsmath, amssymb,color}
\usepackage{hyperref} 
\begin{document}

\title{Violating Lorentz invariance minimally by the emergence of nonmetricity? A Perspective}

\author{Yuri N. \surname{Obukhov}}
\email{obukhov@ibrae.ac.ru} 
\affiliation{Theoretical Physics Laboratory, Nuclear Safety Institute,
Russian Academy of Sciences, B.Tulskaya 52, 115191 Moscow, Russia}

\author{Friedrich W. \surname{Hehl}}
\email{hehl@thp.uni-koeln.de}
\affiliation{Faculty of Mathematics and Natural Sciences, Institute for Theoretical Physics,
University of Cologne, 50923 Cologne, Germany}
\date{Nonmetricity06.tex 03 September 2024}
\begin{abstract}
Lorentz invariance belongs to the fundamental symmetries of nature. It is basic for the successful Standard Model of Particle Physics. Nevertheless, within the last decades, Lorentz invariance has been repeatedly questioned. In fact, there exist different research programs addressing this problem. We argue that a most adequate understanding of a possible violation of Lorentz invariance is achieved in the framework of the gauge-theoretic approach to gravity: a non-vanishing {\it nonmetricity} of a metric-affine geometry of spacetime heralds the violation of the Lorentz symmetry.
\end{abstract}
\maketitle

\section{Euclidean geometry}

Mainly the ancient Greeks constructed a two-dimensional geometrical framework, called nowadays Euclidean geometry. The basic idea was that, say on a table top, if extrapolated in all two-dimensional directions, there are no preferred points and there are no preferred directions. The space is homogeneous and isotropic. In modern parlance, objects in the two-dimensional space, say a material triangle, are invariant under translations and rotations: we call this a two-dimensional Euclidean group of motion.

The third direction, the vertical direction, was different. It was, indeed, a preferred direction: all objects fall to the floor. It was in the time of Newton (17th century), when he and others discovered the gravitational force, which is, as we know, relevant for everyday life and, additionally, for the motion of the heavenly bodies: the Sun, the Moon, and the stars.

In a freely falling elevator or, even better, on the International Space Station ISS, up to minute gravitational {\it tidal}  forces, you don't feel gravitational forces! And, suddenly, the third vertical direction becomes, like the horizontal directions, homogeneous and isotropic. From two-dimensional Euclidean geometry, we arrive at the three-dimensional Euclidean geometry with its group of motion encompassing 3 translations and 3 rotations.

It was the recognition of the properties of the {\it gravitational field} which allowed us to discover the three-dimensional Euclidean geometry as fundamental for the description of nature.

The deep idea to relate the geometric properties of a manifold to the underlying symmetry group goes back to the 19th century Erlangen program of Felix Klein \cite{Erlangen}. Similar and even more general views were advocated by Schouten and van Dantzig \cite{Moskau}, who treated geometry as a game of groups and the conservation of their invariants, described in terms of ``numbers'' or ``objects''.

\section{Gauge theories of internal groups}

In physics, these ideas were eventually reworked by Weyl, Fock and, later, by Yang and Mills \cite{YM} into a gauge theory. The latter is a heuristic field-theoretic formalism developed in the framework of the Lagrange approach in flat spacetime; its purpose is to derive a physical interaction on the basis of a conservation law for the Noether current arising from the rigid symmetry (with constant global transformations) by extending the latter to a local one (with spacetime-dependent transformations), see \cite{Hehl:2020}. 

In high energy particle physics, the Standard Model is constructed as a gauge theory for the internal unitary symmetry groups \cite{ORaifeartaigh:1978jea,TianYuCao}. In this sense, all modern physical theories appear to be based on the gauge approach, with an intriguing exception for the gravitational interaction.

\section{Gauge theories of gravity}

The development of the gauge approach for gravity has a fairly long history. Utiyama \cite{Utiyama:1956} had made a first attempt in this direction by suggesting to consider the Lorentz group $SO(1,3)$ as a gauge symmetry explaining the existence and describing the dynamics of the gravitational field. However, such a scheme fundamentally failed because the corresponding Noether current for the Lorentz group is identified with the angular momentum density, whereas from Newton's theory we know that the matter source of gravity is the mass density, extended to the energy-momentum current of matter  in the special relativistic framework. The latter consistently couples to the group $T(4)$ of spacetime translations.

The 10-parameter {\it Poincar\'e} (or the inhomogeneous Lorentz) group, defined as the semi-direct product $T(4)\rtimes SO(1,3)$ of the 4-parameter group $T(4)$ of translations times the 6-parameter Lorentz group $SO(1,3)$, is the group of motions of the four-dimensional Minkowski spacetime manifold. In other words, the latter is invariant under rigid (`global') Poincar\'e transformations. Wigner \cite{Wigner:1939cj} had established the fundamental classification of quantum mechanical systems with respect to {\it mass and spin} values, labeling the irreducible representations of the Poincar\'e group.

\'Elie Cartan (see \cite[Chap.~2]{Blagojevic:2013}) has been the first one who, in the early 1920s, had, in fact, proposed a construction of the gravity theory as a gauge theory for the Poincar\'e group with mass $m$ and spin $s$ as the two corresponding Noether currents. Subsequently, in the early 1960s, Sciama \cite{Sciama:1962} and Kibble \cite{Kibble:1961} established the simplest gauge-theoretic model of gravity, currently known as the Einstein-Cartan(-Sciama-Kibble) theory, as a {\it viable} generalization of Einstein's general relativity theory that is consistent with experiment. In that approach, the energy-momentum and the spin angular momentum currents arise as the two matter sources of the gravitational gauge field.

\section{Metric-affine gravity (MAG)}

Progress in the construction of Poincar\'e gauge gravity prompted a development of an extension of this gauge-theoretic approach from the Poincar\'e group  $T(4)\rtimes SO(1,3)$ to the four-dimensional general affine group $A(4,R) = T(4)\rtimes GL(4,R)$, see \cite[Chap.\,9]{Blagojevic:2013} and \cite{MAG}. The gauging of the latter symmetry results in a metric-affine spacetime geometry with two independent {\it gravitational potentials}: the metric $g_{ij}$ and the linear connection $\Gamma_{ij}{}^k$. The geometry of the spacetime manifold  is then characterized, in general, by the three nontrivial geometrical objects: the tensors of {\it curvature}
\begin{equation}\label{curv}
R_{kli}{}^j := \partial_k\Gamma_{li}{}^j - \partial_l\Gamma_{ki}{}^j
+ \Gamma_{kn}{}^j \Gamma_{li}{}^n - \Gamma_{ln}{}^j\Gamma_{ki}{}^n,
\end{equation}
{\it torsion}
\begin{equation}\label{tor}
T_{kl}{}^i := \Gamma_{kl}{}^i - \Gamma_{lk}{}^i,
\end{equation}
and {\it nonmetricity}:
\begin{equation}\label{nonmet}
Q_{kij} := -\,\nabla_k g_{ij} = - \,\partial_kg_{ij} + \Gamma_{ki}{}^lg_{lj} + \Gamma_{kj}{}^lg_{il}.
\end{equation}
These three objects are interpreted as the {\it gravitational field strengths} of MAG. 

Let us recall that already in classical mechanics the Galilei-Newton spacetime can be consistently described in terms of a four-dimensional metric-affine geo\-metry encoded in a degenerate metric and a linear connection \cite{Weyl1952,Havas}, for more details see Kopczy\'nski and Trautman \cite[Chap.~3]{Kopczynski}. A clear dichotomy of the spacetime metric $g$ and the linear connection $\Gamma$ is quite typical for the whole of relativistic physics. Einstein definitely was aware of the different physical roles of the metric and the connection structures. As he formulated in one of his last publications \cite{EinsteinPantaleo} (also see \cite[Appendix II]{Meaning}), because of the principle of inertia,  ``...the essential achievement of general relativity, namely to overcome ‘rigid’ space (ie the inertial frame), is {\it only indirectly} connected with the introduction of a Riemannian metric. The directly relevant conceptual element is the ‘displacement field’ ($\Gamma^l_{ik}$), which expresses the infinitesimal displacement of vectors. It is this which replaces the parallelism of spatially arbitrarily separated vectors fixed by the inertial frame (ie the equality of corresponding components) by an infinitesimal operation. This makes it possible to construct tensors by differentiation and hence to dispense with the introduction of ‘rigid’ space (the inertial frame). In the face of this, it seems to be of secondary importance in some sense that some particular $\Gamma$ field can be deduced from a Riemannian metric...''.

In the framework of MAG, the metric tensor $g_{ij}$ and the linear connection $\Gamma_{ki}{}^j$ are considered as  independent fundamental gravitational field variables. Their dynamics is determined by the total Lagrangian of minimally interacting gravitational $(g, \Gamma)$ and matter $\psi^A$ fields: 
\begin{equation}\label{Ltot}
L = V(g_{ij}, R_{ijk}{}^l, T_{ki}{}^j, Q_{kij}) + L_{\rm mat}(g_{ij}, \psi^A, \nabla_i\psi^A).
\end{equation}
The gravitational field Lagrangian $V$ is in general an arbitrary diffeomorphism invariant function of the curvature (\ref{curv}), the torsion (\ref{tor}), and the nonmetricity (\ref{nonmet}). The Lagrangian $L_{\rm mat}$ depends on the matter fields $\psi^A$ (where the condensed index ${}^A$ labels an unspecified number of components, which is irrelevant here) and on their covariant derivatives. The standard set of the field equations of MAG encompasses the so-called ``first'' and ``second'' field equations:
\begin{eqnarray}
{\stackrel * \nabla}{}_lH^{kl}{}_i - E_i{}^k &=& - \Sigma_i{}^k,\label{1st}\\
{\stackrel * \nabla}{}_lH^{kli}{}_j - E^{ki}{}_j &=& \Delta^i{}_j{}^k.\label{2nd}  
\end{eqnarray}
Here the modified covariant derivative ${\stackrel * \nabla}{}_l A^{il\dots}{}_{\dots} = (\nabla_l - T_{kl}{}^k - {\frac 12}Q_{lk}{}^k)A^{il\dots}{}_{\dots} + {\frac 12}T_{mn}{}^i A^{mn\dots}{}_{\dots}$ is introduced for an arbitrary skew-symmetric tensor $A^{ij\dots}{}_{\dots} = -\,A^{ji\dots}{}_{\dots}$, and one defines the field momenta as derivatives of the MAG Lagrangian with respect to the gravitational field strengths (\ref{curv})-(\ref{nonmet}):
\begin{eqnarray}
H^{kli}{}_j := -\,2{\frac {\partial V}{\partial R_{kli}{}^j}},\qquad
H^{ki}{}_j := -\,{\frac {\partial V}{\partial T_{ki}{}^j}},\qquad
M^{kij} := -\,{\frac {\partial V}{\partial Q_{kij}}}.\label{HH}
\end{eqnarray}
Furthermore, from the last two expressions one constructs the gravitational hypermomentum current
\begin{equation}\label{EN}
E^{ki}{}_j := - \,H^{ki}{}_j - M^{ki}{}_j,
\end{equation}
and from all of them the energy-momentum current of the gravitational field:
\begin{equation}\label{Eg}
E_i{}^k := \delta_i^k\,V + {\frac 12}Q_{iln}\,M^{kln} + T_{il}{}^n\,H^{kl}{}_n + R_{iln}{}^m\,H^{kln}{}_m.
\end{equation}
Finally, on the right-hand sides of (\ref{1st}) and (\ref{2nd}) the physical matter sources of the gravitational field are consistently identified with the canonical energy-momentum current of matter and the canonical hypermomentum current of matter, respectively:
\begin{eqnarray}
\Sigma_i{}^k &:=& {\frac {\partial L_{\rm mat}}{\partial\nabla_k\psi^A}}\,\nabla_i\psi^A
- \delta^k_i\,L_{\rm mat},\label{canD}\\
\Delta^i{}_j{}^k &:=& {\frac {\partial L_{\rm mat}}{\partial \Gamma_{ki}{}^j}}.\label{tD}
\end{eqnarray}
The latter expression has a ``fine structure'': We find as its irreducible pieces the {\it canonical currents} of {\it spin} $S_{ij}{}^k := \Delta_{[ij]}{}^k$, of {\it dilation} $J^k := \Delta^i{}_i{}^k$, and of {\it shear} $\Upsilon_{ij}{}^k := \Delta_{(ij)}{}^k - {\frac 14}g_{ij}J^k$. 

The spin current $S_{ij}{}^k$ and the dilation current $J^k$ are well-known from classical field theory. The shear current was introduced in 1976, see \cite{Znat1}. All these currents of spin, dilation, and shear are composed, in an appropriate special-relativistic limit, of an orbital part $x^i\Sigma_j{}^k$ plus an intrinsic part $\Delta^i{}_j{}^k$. Thus, the total hypermomentum:
\begin{equation}
{}^{\rm tot}\Delta^i{}_j{}^k=x^i\Sigma_j{}^k+\Delta^i{}_j{}^k\,,
\end{equation}
for details see the review \cite{Gronwald:1991st}. The intrinsic part of angular momentum has with `spin' an own name. In contrast, for dilation and shear one has to add `total', `orbital' or `intrinsic' in order to specify which part is meant. Conventionally, the nomenclature is as follows: `dilation current' $\longrightarrow$ intrinsic dilation current and `shear current' $\longrightarrow$ intrinsic shear current.

The physical nature of the shear is the subject of the current research, that resulted in the discovery of the new physical concepts of the world spinors and manifields \cite{NS1,NS2,NS3}, which establishes an intriguing bridge between the gravitational physics and the physics of strong interactions.

In general, solving the MAG field equations (\ref{1st}) and (\ref{2nd}) is a fairly difficult problem. However, for the class of models with Lagrangians that are purely quadratic in torsion and nonmetricity, the second equation (\ref{2nd}) reduces to an algebraic relation between $T_{kl}{}^i$, $Q_{kij}$, and the hypermomentum $\Delta^i{}_j{}^k$ of matter, like in the Einstein-Cartan theory. The latter can be solved exactly and then (\ref{1st}) is recast into an effective Einstein field equation \cite{Iosifidis:22}.

The general affine group $A(4,R) = T(4)\rtimes GL(4,R)$ (as well as the Poincar\'e group as a subgroup of the $A(4,R)$) includes the group of translations $T(4)$, that naturally relates any {\it two} neighboring points of the spacetime manifold to each other. When the translational symmetry is violated, one should expect an essential modification of the underlying spacetime geometry.  In particular, a discussion of the ``square roots'' of the 4 translation generators in terms of the 4 additional supersymmetry generators results in the construction of the simple supergravity model, see \cite[Chap.~12]{Blagojevic:2013}. Then advanced geometrical methods are required for the description the spacetime (super)geometry in this case. Here we do not try to break translation invariance, we rather will focus on a possible violation of the Lorentz symmetry.

\section{Violating Lorentz invariance: nonmetricity}

As compared to the group of translations, the Lorentz group $SO(1,3)$ acts locally at {\it one} and only one point in spacetime. As a result, local Lorentz invariance can be formally treated just like an internal symmetry group, such as the $SU(3)$, for example. However, in accordance with the present state of knowledge, Lorentz invariance should be viewed as an external (spacetime) group, which suggests that the violation of the Lorentz symmetry is naturally described by an expansion of the $SO(1,3)$ to the general linear group $GL(4,R)$. From the point of view of physics, a direct manifestation of the violation of the Lorentz invariance is that {\it the light cone looses its status as an absolute element of spacetime}. On the other hand, from the point of view of geometry, a change of the gauge group from the Poincar\'e symmetry to the general affine symmetry results in the emergence of a nontrivial nonmetricity $Q_{kij}\neq 0$ on the spacetime manifold. This leads to an extension of the Riemann-Cartan geometry to a general metric-affine geometry, where the linear connection $\Gamma_{kj}{}^i$ is no longer compatible with the spacetime metric $g_{ij}$. Thus, lengths and angles are not preserved under parallel transport. 

The formalism of the metric-affine spacetime geometry provides a fully consistent description of the possible violation of the Lorentz invariance. In the current literature \cite{Colladay:1998,Jacobson:2007,Kostelecky:2022}, however, the latter is more often discussed by making use of various non-geometric formulations in terms of non-dynamical Lorentz-violating tensor fields (normally arising from nontrivial vacuum expectation values of appropriate quantum operators) constructed from a timelike vector field that has a status of a new {\it absolute element} of the spacetime structure. The naturalness of such non-geometric Lorentz-violating mechanisms is an open issue \cite{Liberati:2013}, though, and the physical essence of an ad hoc Lorentz covariant (!) vector field, responsible for the symmetry breaking, is unclear. Instead of using such artificial constructs for nonminimal violation of the Lorentz symmetry, one should rather turn to the framework of metric-affine gravity where an intrinsic geometric nonmetricity field unmistakably demonstrates the breaking of the Lorentz symmetry. 

One can naturally interpret such a geometric approach as a {\it minimal} violation of the Lorentz symmetry. The most spectacular physical manifestation of the latter would be the destruction of the light cone that is no longer invariant under the general affine symmetry group. We described this phenomenon as a short range quantum mechanical ``smearing out'' of the light cone in \cite{Znat3}.

The question of probing the nonmetricity of spacetime actually amounts to the most fundamental aspects of establishing the geometric structure of spacetime in an operational way directly from physical observations by making use of the suitable measuring tools, such as clocks, rigid bodies, light rays, freely falling test particles, etc. Such a peculiar feature of nonmetricity as the {\it non-integrability of length} was discussed in detail already by Eddington \cite[Sec.~84]{Eddington} in relation to the early theory of Weyl \cite{Weyl1952,Hehl:1988,Scholz:2018}.

An ambitious program of deriving the spacetime structure from the free fall dynamics of test particles and light propagation was put forward by Ehlers, Pirani and Schild (EPS) \cite{EPS1,EPS2,EPS3}. Among other findings, the EPS approach proposed the {\it second clock effect} to characterize the manifestations of a nontrivial nonmetricity. In simple terms, one takes two identical clocks completely synchronized at an initial spacetime point $P_1$ and transfers them to a final point $P_2$ along two different paths. As compared to the usual relativistic (``first clock'') effect, manifested in the different elapsed times  measured at paths' end, the ``second clock effect'' predicts different rates with which clocks tick at $P_2$.

Although the original EPS discussion \cite{EPS1,EPS2} was focused on the particular case of Weyl geometry, the non-integrability of length and the second clock effect are inherent to a general spacetime nonmetricity. When special relativity is valid locally, the second clock effect (or the twin paradox of second kind) should be absent.

\section{Hypermomentum}

In contrast to the possibility of testing the Riemannian spacetime geometry with the help of structureless point particles in Einstein's general relativity, one needs to use the matter with microstructure to probe the post-Riemannian spacetime structure in the framework of the metric-affine approach with independent metric and connection. Einstein \cite{Einstein:2002} stressed that ``...the question whether this continuum has a Euclidean, Riemannian, or any other structure is a question of physics proper which must be answered by experience, and not a question of a convention to be chosen on grounds of mere expediency.'' Experimental investigation of the spacetime structure requires the analysis of the corresponding equations of motion of extended text bodies propagating in generalized metric-affine geometries. Quite importantly, these equations of motion are derived from first principles on the basis of conservation laws for dynamical currents, and they clearly demonstrate that the nonmetricity is coupled only to the intrinsic microstructure of a test particle and there is not any interaction with the orbital hypermomentum of extended material bodies \cite{PO:2007,PO:2014,Iosifidis}.

A viable description for the physical matter with microstructure is provided by the hyperfluid model \cite{hyper1,hyper2,Aldrovandi}, a classical continuous medium characterized by the pressure $p$ and an average 4-velocity $u^i$; the corresponding matter sources (\ref{canD}) and (\ref{tD}) read, respectively:
\begin{eqnarray}
\Sigma_i{}^k &=& {\cal P}_i\,u^k - p\left(\delta^k_i - {\frac {u^ku_i}{c^2}}\right),\label{EF}\\
\Delta^i{}_j{}^k &=& {\cal J}^i{}_j\,u^k\,.\label{DF}
\end{eqnarray}
Elements of such a fluid carry the 4-momentum density ${\cal P}_i$ and the intrinsic hypermomentum density ${\cal J}^i{}_j$. 

The bottom line looks as follows: When the Lorentz invariance in broken and the light cone looses its status of an absolute element, the decisive physical observation should become the detection of a nontrivial {\it nonmetricity} of spacetime.

\end{document}